\documentclass[reprint, amsmath, amssymb, aps, nofootinbib, superscriptaddress, longbibliography]{revtex4-1} %DO NOT CHANGE THIS
\usepackage{multirow}
\usepackage{verbatim}
\usepackage{xfrac}
\usepackage{graphicx}% Include figure files
\usepackage{dcolumn}% Align table columns on decimal point
\usepackage{bm}% bold math
\usepackage{hyperref}% add hypertext capabilities
\usepackage{url}  %Required
\usepackage{booktabs}
\usepackage{dcolumn}
\usepackage{tabularx}
\usepackage{xcolor,listings}
\lstdefinestyle{custompython}{
  belowcaptionskip=1\baselineskip,
  breaklines=true,
  frame=L,
  xleftmargin=\parindent,
  language=Python,
  showstringspaces=false,
  basicstyle=\footnotesize\ttfamily,
}

\begin{document}

\author{Allison C. Morgan}
 \email{allison.morgan@colorado.edu}
 \affiliation{Department of Computer Science, University of Colorado, Boulder, CO, USA}
\author{Samuel F. Way}
 \email{samuel.way@colorado.edu}
 \affiliation{Department of Computer Science, University of Colorado, Boulder, CO, USA}
\author{Aaron Clauset}
 \email{aaron.clauset@colorado.edu}
 \affiliation{Department of Computer Science, University of Colorado, Boulder, CO, USA}
 \affiliation{BioFrontiers Institute, University of Colorado, Boulder, CO, USA}
 \affiliation{Santa Fe Institute, Santa Fe, NM, USA}

\title{Automatically assembling a full census of an academic field}

\begin{abstract}
The composition of the scientific workforce shapes the direction of scientific research, directly through the selection of questions to investigate, and indirectly through its influence on the training of future scientists. In most fields, however, complete census information is difficult to obtain, complicating efforts to study workforce dynamics and the effects of policy. This is particularly true in computer science, which lacks a single, all-encompassing directory or professional organization. A full census of computer science would serve many purposes, not the least of which is a better understanding of the trends and causes of unequal representation in computing. Previous academic census efforts have relied on narrow or biased samples, or on professional society membership rolls. A full census can be constructed directly from online departmental faculty directories, but doing so by hand is prohibitively expensive and time-consuming. Here, we introduce a topical web crawler for automating the collection of faculty information from web-based department rosters, and demonstrate the resulting system on the 205 PhD-granting computer science departments in the U.S. and Canada. This method constructs a complete census of the field within a few minutes, and achieves over 99\% precision and recall. We conclude by comparing the resulting 2017 census to a hand-curated 2011 census to quantify turnover and retention in computer science, in general and for female faculty in particular, demonstrating the types of analysis made possible by automated census construction.
\end{abstract}

\maketitle

\section{Introduction}
\label{sec:introduction}

Tenured and tenure-track university faculty play a special role in determining the speed and direction of scientific progress, both directly through their research and indirectly through their training of new researchers. Past studies establish that each of these efforts is strongly and positively influenced through various forms of faculty diversity, including ethnic, racial, and gender diversity. 
As an example, research shows that greater diversity within a community or group can lead to improved critical thinking~\cite{vandyne1996conflict} and more creative solutions to complex tasks~\cite{mcleod1996ethnic,page2008} by pairing together individuals with unique skillsets and perspectives that complement and often augment the abilities of their peers. 
Additionally, diversity has been shown to produce more supportive social climates and effective learning environments~\cite{milem2003educational}, which can facilitate the mentoring of young scientists. Despite these positive effects, however, quantifying the impact of diversity in science remains exceedingly difficult, due in large part to a lack of comprehensive data about the scientific workforce.

Measuring the composition and dynamics of a scientific workforce, particularly in a rapidly expanding field like computer science, is a crucial first step toward understanding how scholarly research is conducted and how it might be enhanced. For many scientific fields, though, there is no central listing of all tenure-track faculty, making it difficult to define a rigorous sample frame for analysis. Further, rates of adoption of services like GoogleScholar and ResearchGate vary within, and across disciplines. For instance, gender representation in computing is an important issue with broad implications~\cite{hill2010so}, but without a full census of computing faculty, the degree of inequality and its possible sources are difficult to establish~\cite{way:gender}. Some disciplines, like political science, are organized around a single professional society, whose membership roll approximates a full census~\cite{fowler2007social}. Most fields, on the other hand, including computer science, lack a single all-encompassing organization and membership information is instead distributed across many disjoint lists, such as web-based faculty directories for individual departments. 

Because assembling such a full census is difficult, past studies have tended to avoid this task
%exhaustively collecting and aggregating these lists, and have 
and have instead used samples of researchers~\cite{cole1979age,allison:departmenteffects,long1995scientific,xie98}, usually specific to a particular field~\cite{myers2011mathematical,amir2008ranking,katz2011reproduction,schmidt2007ranking,hanneman2001prestige}, and often on the scientific elite~\cite{zuckerman1977scientific,schlagberger2016institutions}. Although useful,
%While it can be valuable to examine unusually successful careers, 
such samples are %generally 
not representative of the scientific workforce as a whole and thus have limited generalizability. 
%More widely applicable results could be obtained by considering the entire workforce rather than a potentially biased subset. 
One of the largest census efforts to date assembled, by hand, a nearly complete record of three academic fields: computer science, history, and business~\cite{clauset:hiring}. This data set has %been useful for 
shed considerable light on dramatic inequalities in faculty training, placement, and scholarly productivity~\cite{clauset:hiring,way:gender,way:misleading}. But, %While comprehensive, 
this data set is only a single snapshot of an evolving and expanding system and hence offers few insights into the changing composition and diversity trends within these academic fields.
%dynamics of the academy, and  and how it reached its current status.

In some fields, yearly data on faculty numbers and composition are available in aggregate. In computer science, the Computing Research Association (CRA) documents trends in the employment of PhD recipients through the annual Taulbee survey of computing departments in North America.\footnote{See {\scriptsize \url{cra.org/resources/taulbee-survey/}} } Such surveys can provide valuable insight into trends and summary statistics on the scientific workforce but suffer from two key weaknesses. First, surveys are subject to variable response rates and the misinterpretation of questions or sample frames, which can inject bias in fine-grained analyses~\cite{groves2011survey,weisberg2009total}. Second, aggregate information provides only a high-level view of a field, which can make it difficult to investigate causality~\cite{imai2011unpacking}. For example, differences in recruitment and retention strategies across departments will be washed out by averaging, thereby masking any insights into the efficacy of individual strategies and policies.

% ----- FIGURE -----
\begin{figure*}
\includegraphics[width=\textwidth]{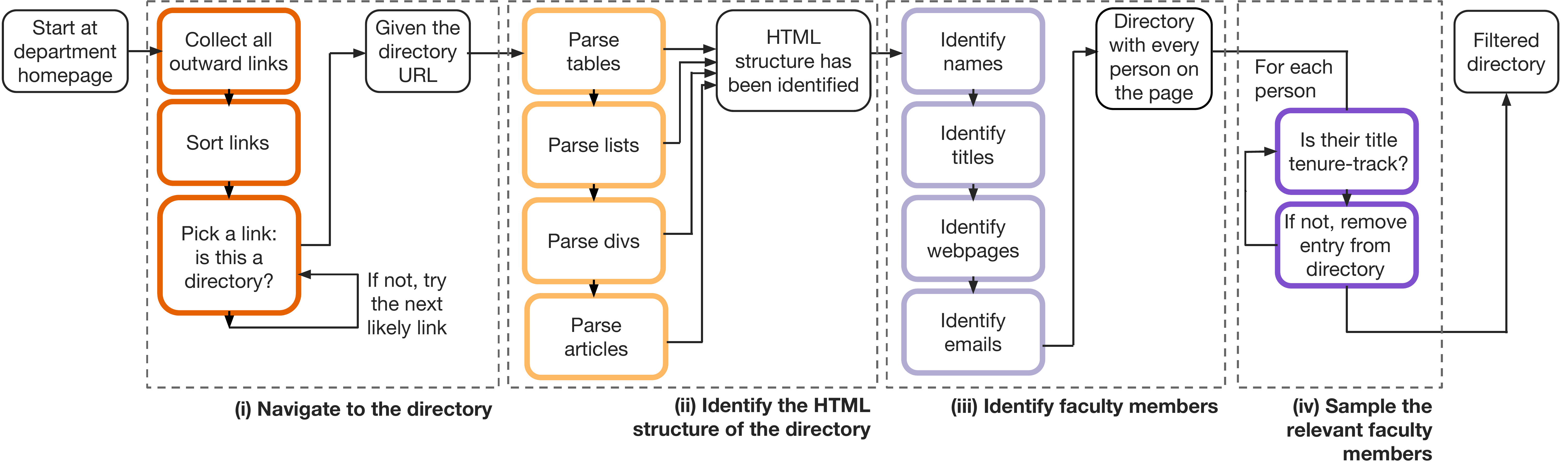}
\caption{{\bf General schematic of our solution to the academic census problem.} {\normalfont Starting from a department's homepage, our web crawler builds a census of its faculty in the following steps: (i) navigate to the department's faculty directory page, (ii) identify the logical structure of the directory, (iii) parse the directory to resolve potential faculty members, and finally (iv) sample and return a list of the relevant faculty members.}}
\label{fig:schematic}
\end{figure*}
% --------------------

Here, we present a novel system, based on a topical web crawler, that can quickly and automatically assemble a full census of an academic field using digital data available on the public World Wide Web. This system is efficient and accurate, and it can be adapted to any academic discipline and used for continuous collection. The system is capable of collecting census data for an entire academic field in just a few hours using off-the-shelf computing hardware, a vast improvement over the roughly 1600 hours required to do this task by hand~\cite{clauset:hiring}. By assembling an accurate census of an entire field from online information alone, this system will facilitate new research on the composition of academic fields by providing access to complete faculty listings, without having to rely on surveys or professional societies. This system can also be used longitudinally to study how the workforce's composition changes over time, which is particularly valuable for evaluating the effectiveness of policies meant to broaden participation or improve retention of faculty. Finally, applied to many academic fields in parallel, the system can elucidate scientists' movement between different disciplines and relate those labor flows to scientific advances. In short, many important research questions will benefit from the availability of accurate and frequently-recollected census data.

Our study is organized as follows. We begin by detailing the design and implementation of our web crawler framework. Next, we present the results of our work in two sections. The first demonstrates the validity and utility of the crawler by collecting census data for the field of computer science and comparing it to a hand-curated census, collected in 2011~\cite{clauset:hiring}. The second provides an example of the type of research enabled by our system and uses the 2011 and 2017 censuses to investigate the ``leaky pipeline" problem in faculty retention.

\section{Background}
\label{sec:background}

Comprehensive data about academic faculty can be compiled from web-based sources, but is widely distributed and inconsistently structured across computer science departments. Here, we introduce a topical web crawler to retrieve and assemble these data into a comprehensive census. As a method for distributed information discovery, a topical web crawler navigates the Web, searching for relevant documents~\cite{menczer:arachnid}. A crawler's search can be broad, such as the Never-Ending Language Learner, which continuously crawls the Web to learn new properties and relationships among persons, places or things~\cite{mitchell:nell}. Or, it can be narrowly focused, such as for building domain-specific Web portals~\cite{mccallum:automating}. Our crawler falls into the latter category in that its search space is restricted to academic webpages, with a goal of navigating to and extracting information from faculty directories.

Our search algorithm is an adapation of a ``best-first search''~\cite{cho:efficient,menczer:topical,pant:crawling,chakrabarti:focused,menczer:evaluating} and can be described as follows: the crawler starts from a department's homepage, and scores each outgoing hyperlink to estimate the probability that it leads to the corresponding department's faculty directory. Then, the crawler visits the links in a greedy order, based on their computed score. If the visited page is not a directory, any additional links found on that page are scored and added to the existing priority queue. Once the topical crawler encounters the faculty directory, it follows the task of extracting the desired information from the page. Like the link structure leading to the page, faculty directories lack a common markup language~\cite{butt:taxonomy} and are instead formatted in a variety of ways. Our method for extracting faculty information from directories therefore must thus be robust and adaptive. We describe our approach in following sections.

\section{Problem Formalization}
\label{sec:methods}

We define the field of computer science to be the 205 North American, PhD-granting institutions from the CRA's Forsythe List.\footnote{See {\scriptsize \url{http://archive.cra.org/reports/forsythe.html}} } Here, the input to our system is a list of department homepages corresponding to each of these institutions, however, we note that searches can proceed from any listing of department or university homepages. Faculty employment information for these universities is contained in web-based faculty directories maintained by each department, yet assembling data from all institutions into a combined census is a difficult and prohibitively expensive task. Crowdsourcing census construction, too, is complicated by the fact that domain expertise is required to distinguish tenured or tenure-track (TTT) from other faculty or staff positions, and for separating computer science (CS) from electrical engineering (EE) or computer engineering (CE) faculty. For example, the title of ``associate professor" generally implies a full-time, tenured position, while an ``adjunct associate professor" is neither full-time, nor tenured. To make such distinctions, workers must receive specific training when collecting data by hand, increasing both the cost and duration of a survey. The 2011 census~\cite{clauset:hiring} took a trained pool of workers about 1600 hours and cost \$16,000. Hence, to generate regular census snapshots, for multiple disciplines, would be prohibitively expensive and require a dedicated, trained workforce. Our topical web-crawler provides a cheap, accurate, and scalable alternative.

The crawler simplifies the overall task by finding and parsing departmental directories in four steps: (i) efficiently navigate to a department's directory, (ii) identify the HTML structure separating entries within the directory, (iii) extract every faculty record by identifying names, titles, webpages, and email addresses, and (iv) filter this list to include only TTT faculty members (Fig.~\ref{fig:schematic}). In steps (i) and (iv), our approach favors higher recall by preferring false positive errors, since false negatives imply either the missed opportunity to scrape a directory or the omission of a TTT faculty member. In this setting, false positives can be corrected via downstream analyses, typically at the cost of extra computation with the parsing of a candidate's resume or the manual verification of very specific information using services like Amazon's Mechanical Turk\footnote{See {\scriptsize \url{www.mturk.com/}} } or CrowdFlower.\footnote{See {\scriptsize \url{www.crowdflower.com/}}} In the following sections, we discuss each of the outlined tasks in the order of their completion.

\medskip 
\paragraph{Navigate to the directory.}

\begin{figure}[t!]
\includegraphics[width=\columnwidth]{{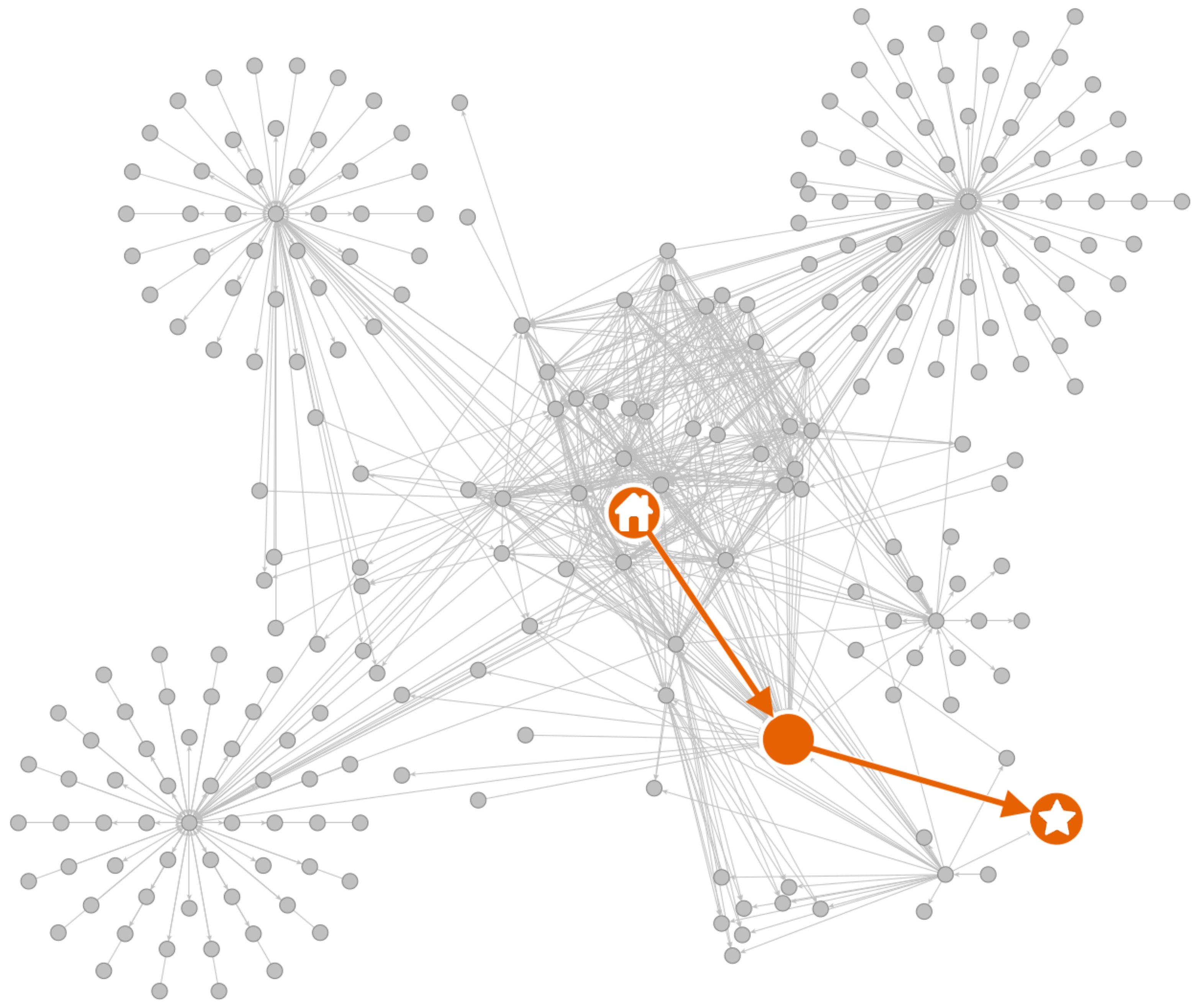}}
\caption{{\bf Example hyperlink network surrounding a department homepage.} {\normalfont The network of all reachable webpages within two hops from the Department of Computer Science at University of California, Davis homepage (home icon). Shown in orange is the shortest possible path---and the one our crawler takes---to reach the targeted faculty directory (star icon).}}

%Some departments have a link to their directory from the homepage, but some do not. For example, shown here is the complete hyperlink network within two hops of the homepage of the Department of Computer Science at University of California, Davis. In orange is the shortest path necessary (and the path our crawler takes) to reach the department's directory.}
\label{fig:hyperlink}
\end{figure}

Our crawler's navigation strategy has two primary components: (i) navigate efficiently from a department's homepage to their directory, and (ii) identify whether a page appears to be a directory. First, in order to navigate to the desired faculty directories, our crawler must decide which hyperlinks to follow. Starting from a department's homepage, it adds all outgoing hyperlinks to a max-priority queue, with priorities set equal to the number of keywords found within each URL and its surrounding text. This keyword list has 25 words, including ``faculty,'' ``directory,'' and ``people,'' which were manually extracted from common features of departments' directory URLs. The crawler then visits pages in order of their priority, keeping track of any URLs that have already been visited, and adding newly discovered URLs to the queue as it goes, until it reaches a directory page (Fig.~\ref{fig:hyperlink}).

For each visited page, the crawler must decide whether it is a directory to parse. To avoid parsing every likely page for faculty members, the crawler uses a random forest classifier to decide whether a page is likely to be worth fully parsing for faculty listing information. Each page is characterized by counting motifs commonly found on faculty directories, such as names, phone numbers, email addresses, and job titles. Since faculty directories typically contain little other text, a page's feature set includes counts of these motifs as a fraction of all words. A false negative, overlooking a faculty directory, is an unrecoverable error, and induces a group of correlated false negatives for faculty in the census. We prefer a directory classifier that has no false negatives at the expense of more false positives, so any pages that yield a likelihood greater than zero are passed to the next stage (see below). Additionally, parsing a non-directory page is relatively inexpensive in terms of computational time and, since no faculty will likely be extracted from this page, these pages are easy to subsequently identify as false positives.

% ----- TABLE -----
%\begin{table}[b!]
%\begin{tabular}{p{1cm}|p{1cm}p{1cm}p{1cm}p{1cm}}
%total & divs & tables & lists & articles \\
%\hline
%205 & 100 & 80 & 24 & 1 \\
%\end{tabular}
%\caption{Counts of faculty directory HTML formats among PhD-granting CS departments in our sample \allie{This table could be cut and the information moved into text above.}} \label{tab:table1}
%\end{table}
% --------------------

% ----- FIGURE -----

\smallskip 
\paragraph{Identify the HTML structure of the directory.}

% --------------------

% ----- FIGURE -----
\begin{figure*}[t!]
\includegraphics[width=\textwidth]{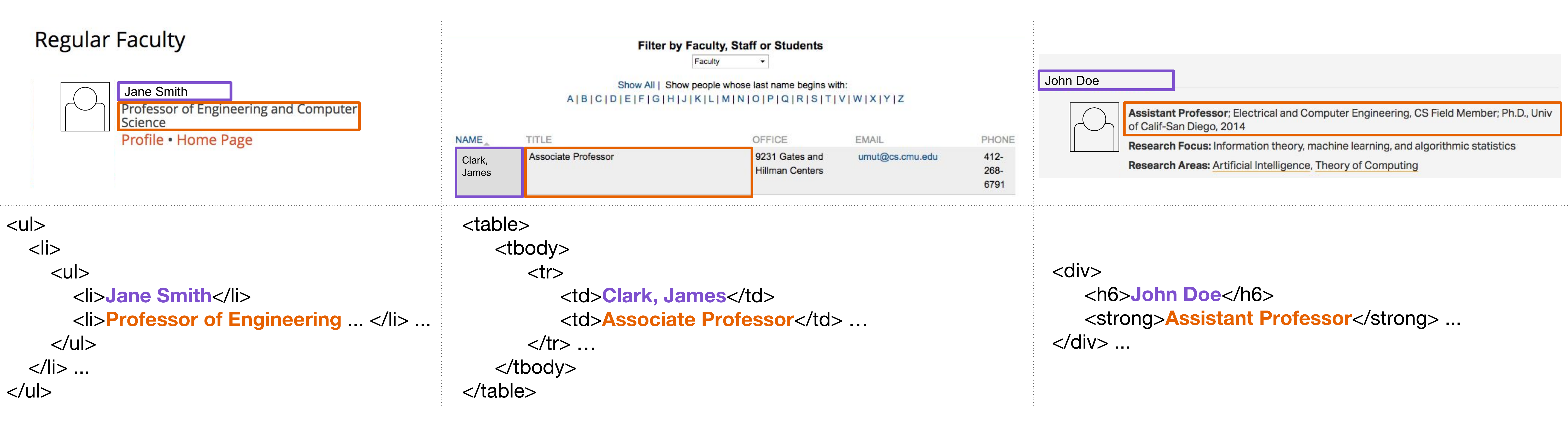}
\caption{{\bf Faculty directories are formatted in a wide variety of styles, but using common structural elements.} {\normalfont Three real examples of directories formatted using lists (left), tables (center), and divs (right). Highlighted are the pieces of information extracted by our crawler from these pages: faculty names (purple) and titles (orange).}}
%\caption{Three real examples displaying the wide variety of HTML structures that may separate faculty information. Shown is the rendered HTML and underlying code for directories using lists, tables and divs. Highlighted are pieces of information extracted from these pages: faculty member's names (purple) and titles (orange).}
\label{fig:html_types}
\end{figure*}
% --------------------

Once the crawler discovers a directory, it must extract information from a variety of HTML formatting conventions (Fig.~\ref{fig:html_types}). In practice, despite enormous variation in the visual styling of these pages, there is a short list of common HTML tags that separate faculty members from each other: divs, tables, lists and articles. %To parse data from these structures, we built four corresponding parsers, one for each HTML format. Our crawler applies each of these parsers to a department's directory and assigns an HTML-format type based on which parser returned the most faculty members. %(Table~\ref{tab:table1}). 
These four structural tags are used to format repeated faculty entries within a directory.
Our crawler attempts to segment a directory according to each of these %four structural
tags, separately, and ultimately selects the segmentation resolving the largest number of faculty records.
Following this procedure for each of the 205 CS departments, we found that 100 directories were formatted with divs, 80 with tables, 24 with lists, and 1 with articles.
%
%When the crawler reaches a webpage that the directory classifier decides is a directory, it applies each of the parsing methods in a sequence (tables, lists, divs, articles). The sequence is stopped early if a method-dependent number of faculty records are obtained; the precise stopping values were determined by hand. 
%In this step, the crawler also identifies whether the department has a multi-page directory. If so, the crawler collects the list of links and applies a parser to each. If no faculty members are collected from the page, the crawler logs the output and moves to the next highest priority URL in the queue.
Finally, as part of this step, the crawler detects whether the faculty directory is distributed across multiple pages by searching for div or list tags containing common ``pagination'' or ``pager'' classes. If detected, the crawler collects the list of links and applies a parser to each. If no faculty members are collected from the page, the crawler logs the output and moves to the next highest priority URL in the queue.

\smallskip 
\paragraph{Identify faculty members.}

After identifying the HTML structure separating faculty members from each other, the next step is to extract faculty information from the page. Each directory consists of repeated HTML elements, and all faculty directories contain similar information: first and last names, titles, email addresses, and faculty homepage links (Fig.~\ref{fig:html_types}). This repetition in HTML and content allows the crawler to distinguish individual faculty records, and extract the target information. Our approach to identifying each faculty attribute is based on a set of keyword-matching heuristics, each based on a whitelist of known relevant strings.

To detect and extract names, we constructed a whitelist of first and last names from the 2011 computer science faculty census~\cite{clauset:hiring}. If a string contained a single substring that can be found in this set of names, the crawler classifies that string as a name. This set contained 6,798 entries. As directories were scraped, they were manually inspected and any previously unseen names were added to our list. This procedure added 259 new names (4\%) to the whitelist. A similar, more exhaustive list of names could be constructed from other publicly available data, e.g., family names from the U.S. Census%
\footnote{See {\scriptsize \url{census.gov/topics/population/genealogy/data.html}}} or author names in bibliographic databases like DBLP.%
\footnote{See {\scriptsize \url{dblp.uni-trier.de/}} }

The crawler then extracts appointment titles and email addresses from the text between names. For titles, we employ a whitelist comprising the set of all conventional titles for TTT and non-TTT faculty using partial string matches. This list is intentionally large, such that we avoid misidentifying a faculty member's title. If the crawler cannot find a title relevant to a name, it omits that entry from the directory. Typically email addresses can be identified using simple regular expressions. In some cases, emails are obfuscated on a directory page; however, in most cases circumventing such efforts is trivial. The most common obfuscation method is to remove any shared suffix (``@colorado.edu''). In these cases (4.9\% of all CS departments), the domain can be trivially inferred from the web domain in the directory URL. Faculty email addresses could not be identified in this way for only 21.5\% of departments.

Lastly, as it is often available, the crawler also searches for faculty webpage URLs. Currently, only webpages included as links surrounding faculty member names are identified. Although they are not utilized in this work, these URLs could be used as input for subsequent collection of faculty curriculum vitas, a direction we leave for future work. 

At the end of this stage, the crawler has derived an exhaustive list of every person on the directory. This list will contain true positives, the records of TTT faculty, as well as false positives, which are any other individuals. This set of records is a superset of the in-sample faculty we seek. The next stage is to remove these false positives.

\smallskip 
\paragraph{Sample the relevant faculty members.} \label{sec:sample}

In addition to TTT faculty, department faculty rosters often list many other kinds of individuals, including affiliated, courtesy, teaching or research faculty, various staff or non-faculty administrative positions, and sometimes trainees like postdocs or graduate students. In Section~\ref{sec:science}, we focus our analysis on TTT faculty for direct comparison to the 2011 census, and hence here we discuss selecting out TTT individuals. This filtering criteria reflects a choice; another filtering criteria could be applied here to produce a different kind of directory, e.g., all research faculty, contingent faculty (adjunct, adjoint, etc.), or teaching faculty.  Another choice we made is the restriction to faculty whose primary affiliation is within CS, which excludes affiliated faculty and courtesy appointments.

To perform this filtering, we construct a blacklist of titles that signify non-TTT faculty and staff (such as ``adjoint,'' ``staff,'' ``emeritus,'' and ``lecturer''). This list contains 84 titles and was constructed by the manual evaluation of faculty records. Faculty records containing these restricted titles are removed from the output directory. Often universities publish online their definitions of non-TTT appointments.%
\footnote{E.g., {\scriptsize \url{faculty.umd.edu/policies/ten_titles.html}} or \\ {\scriptsize \url{ap.washington.edu/ahr/academic-titles-ranks/}}}
A more sophisticated approach might collect these documents to build department specific filters. 

Some CS faculty are housed in joint Electrical Engineering and Computer Science (EECS) departments, and so the crawler also checks whether a person is flagged as computer science faculty. For example, if a title contains the substrings ``of,'' ``from,'' or ``in,'' it checks whether that string contains a computing related word from a short custom built whitelist. However, in most cases, information about which field, CS or EE, a faculty member officially belongs to is not available on the directory. We address this issue manually in Sec.~\ref{sec:science}. Previous work has shown that faculty research interests can be distinguished using topic modeling on publication titles~\cite{way:gender}. In the future, filtering faculty by research field in this stage could potentially be automated using publication data.

\section{Results}
\label{sec:results}

The modular design of our system allows us to evaluate both how individual stages behave independently of each other and collectively. First, we evaluate each of the four stages separately, discussing errors and where future work could improve the system's behavior. Then, we analyze their combination as a single system. All evaluations of the timing of our system have been made with any HTML already requested and stored locally, which controls for variability due to network latency and server liveness. Finally, we assess the generality of the system by deploying it on two additional fields, noting potential improvements for further expansion.

\subsection{Navigate and classify}
\label{sec:navigating}

\begin{figure}[t!]
\includegraphics[width=\columnwidth]{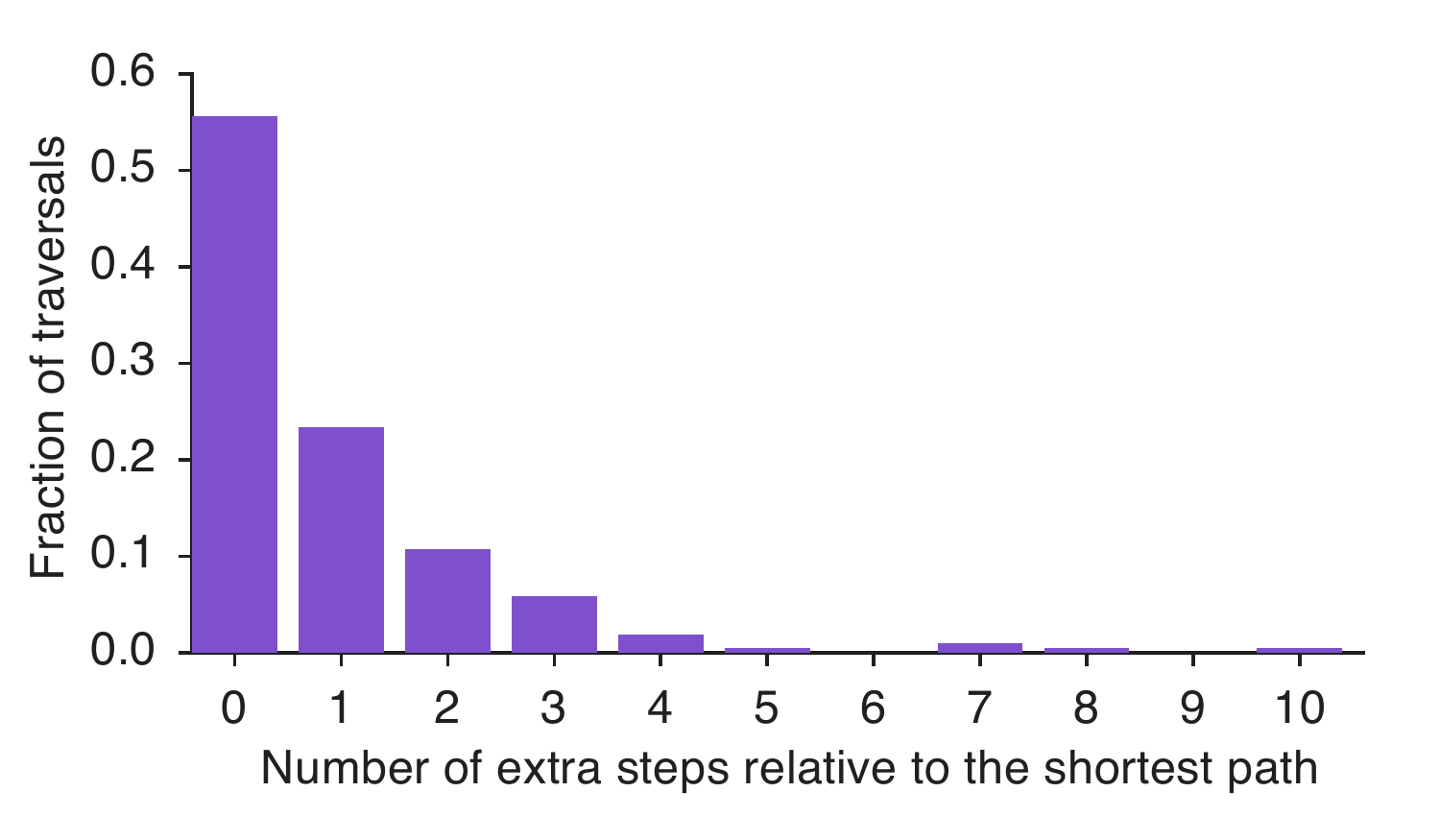}
\caption{{\bf The distribution of extra steps taken in navigating to faculty directories.} {\normalfont The number of steps taken by the crawler, subtracting the minimum path length from each department's homepage to the corresponding faculty directory. In 79\% of cases, our crawler commits only 1 extra step beyond the optimal path length.}}
%\caption{Distribution of the number of steps taken by our crawler minus the shortest possible path length between the department homepage and the department's faculty directory. In 79\% of cases, our crawler is within 1 extra step of the optimal path length.}
\label{fig:bfs}
\end{figure}

We evaluate the efficiency of our crawler's navigation strategy by comparing its traversal to the shortest path from the homepage to the directory (Fig.~\ref{fig:bfs}). A difference of zero means that our crawler makes as many HTTP requests as the shortest path. For more than half (56\%) of departments, our navigation heuristic is optimal, and on average makes only 0.88 excess HTTP requests relative to optimality.

% ----- FIGURE -----
% --------------------
%
Next, to evaluate the performance of our directory classifier, we run a stratified five-fold cross-validation test. The positive training set consists of all 205 department directories, and the negative training set contains a uniformly random 50\% sample (4206) of non-directory pages linked from the department homepages. As suggested above, the crawler was designed to avoid false negatives. In this case, a false negative would cause the crawler to not parse a directory and therefore induce a group of correlated false negatives in the census. To reduce this likelihood, the classifier returns a positive if the directory likelihood for a page is greater than zero. 
The resulting classifier has perfect recall across all five folds, at the expense of precision, as intended. The average accuracy---fraction of correct classifications (positive and negative)---is 0.82 due to the over-classification of non-directory webpages as faculty directory pages (standard deviation of 0.02), and the average area under the ROC curve is 0.99. The non-directory pages that are particularly difficult for the classifier to distinguish are primarily pages listing campus or administrative contact information. These pages often have similar features to directories (names, phone numbers and email addresses) and little other text. For similar reasons, pages that contain job postings or directories of affiliated or courtesy faculty are also flagged as directories. %Since our design is modular, false positives at one stage can be corrected in a subsequent stage when more expensive analysis can be applied. 
False positives produced here are largely filtered out as non-TTT faculty in fourth and final stage, as described below.

Combining efficient navigation and directory classification yields a considerable improvement over a naive breadth-first search. The average time to parse a page is 24 CPU seconds and a breadth-first crawl visits 62 pages, on average, to find the directory. Thus, the most naive implementation of a crawler would take about 1488 CPU seconds per department. In comparison, we find empirically that the navigation approach detailed here, without the directory classifier, takes 57 CPU seconds on average, while navigating intelligently and using the directory classifier takes only 55 CPU seconds.

\subsection{Parse and filter}
\label{sec:sample}
We evaluate the performance of our four parsing methods and our ability to recover the correct attributes of a faculty record, by manually verifying their output on a subset of departments. This subset is composed of 69 departments, chosen uniformly at random but conditioned on having at least one representative for each of the four HTML structures. %This resulted in a sample consisting of 69 departments. 
To each of these departments, we apply the correct parser directly to its faculty directory and inspect the results by hand.

To evaluate our parsing method's accuracy, precision and recall are measured by manually counting the number of TTT faculty. The 69 directories in our evaluation group list 1872 TTT faculty, of which 1868 are correctly identified, leaving 4 members missing due to either ill-formatted HTML or a missing title. The parsers also misclassifiy 12 individuals, calling them TTT computer science faculty when they are actually emeritus, affiliated faculty, or staff. 

On this sample, the parser's recall is 99.97\%, indicating that only a small fraction of true TTT faculty are missed. And, the system's precision is 99.36\%, indicating that only a small fraction of non-TTT faculty are incorrectly included. The directory parsing stage is the most time intensive step of our system, taking on average 47 CPU seconds per department. As we will discuss in Sec.~\ref{sec:deploying}, this is a dramatic improvement over previous work.

% ----- TABLE -----
%\begin{comment}
\begin{table*}
\begin{tabular}{rrrrrrrrrrrrrrrrrrr}
                          &                           & & \multicolumn{16}{c}{MTurk}                                                                                                                                                                                                                                                                                                  \\
                          &                           & $n$ & \rotatebox{90}{New $\rightarrow$ Asst} & \rotatebox{90}{New $\rightarrow$ Assoc} & \rotatebox{90}{New $\rightarrow$Full} & \rotatebox{90}{New $\rightarrow$ Gone} & \rotatebox{90}{Asst $\rightarrow$ Asst} & \rotatebox{90}{Asst $\rightarrow$ Assoc} & \rotatebox{90}{Asst $\rightarrow$Full} & \rotatebox{90}{Asst $\rightarrow$ Gone} & \rotatebox{90}{Assoc $\rightarrow$ Asst} & \rotatebox{90}{Assoc $\rightarrow$ Assoc} & \rotatebox{90}{Assoc $\rightarrow$ Full} & \rotatebox{90}{Assoc $\rightarrow$ Gone} & \rotatebox{90}{Full $\rightarrow$ Asst} & \rotatebox{90}{Full $\rightarrow$ Assoc} & \rotatebox{90}{Full $\rightarrow$ Full} & \rotatebox{90}{Full $\rightarrow$ Gone} \\ \cline{4-7}
\multirow{15}{*}{\rotatebox{90}{Crawler}} & New $\rightarrow$ Asst   & 89 & \multicolumn{1}{|r}{94.4} & 0.0 & 0.0 & \multicolumn{1}{r|}{4.5} & 1.1 &  &  &  &  &  &  &  &  &  &  &  \\
                          & New $\rightarrow$ Assoc  &17 & \multicolumn{1}{|r}{5.9} &  52.9 & 0.0 &  \multicolumn{1}{r|}{17.6} &  &  &  &  &  & 23.5 &  &  &  &  &  &  \\
                          & New $\rightarrow$ Full  & 26 & \multicolumn{1}{|r}{0.0} & 3.8 & 46.2 & \multicolumn{1}{r|}{26.9} &  &  &  &  &  &  &  &  &  &  & 23.1 &  \\ \cline{4-7} \cline{8-11}
                          & Asst $\rightarrow$ Asst  & 14 & & & & &  \multicolumn{1}{|r}{50.0}  & 50.0 & 0.0 & \multicolumn{1}{r|}{0.0} & &  &  &  &  &  &  &  \\
                          & Asst $\rightarrow$ Assoc  & 35 & & & & & \multicolumn{1}{|r}{0.0} & 91.4 & 8.6 & \multicolumn{1}{r|}{0.0} & &  &  &  &  &  &  &  \\
                          & Asst $\rightarrow$ Full    & 6 & &  & &  & \multicolumn{1}{|r}{16.7} &  50.0 & 33.3 & \multicolumn{1}{r|}{0.0} &  &  &  &  &  &  &  &  \\
                          & Asst $\rightarrow$ Gone   & 24 & & & & & \multicolumn{1}{|r}{16.7} & 33.3 & 8.3 & \multicolumn{1}{r|}{41.7} &  &  &  &  &  &  &  &  \\ \cline{8-11} \cline{12-15}
                          & Assoc $\rightarrow$ Asst  & 3 & & & & & & & & & \multicolumn{1}{|r}{33.3} & 33.3 & 0.0 & \multicolumn{1}{r|}{33.3} &  &  &  &  \\
                          & Assoc $\rightarrow$ Assoc & 65 & & & & & & & & & \multicolumn{1}{|r}{0.0} & 100.0 & 0.0 & \multicolumn{1}{r|}{0.0} &  &  &  &  \\
                          & Assoc $\rightarrow$ Full  & 53 & & & & & & & & & \multicolumn{1}{|r}{0.0} & 1.9 & 98.1 & \multicolumn{1}{r|}{0.0} & &  &  &  \\
                          & Assoc $\rightarrow$ Gone & 45 & & & & & & & & & \multicolumn{1}{|r}{0.0} & 44.4 & 13.3 & \multicolumn{1}{r|}{42.2} &  &  &  &  \\ \cline{12-15}\cline{16-19}
                          & Full $\rightarrow$ Asst    & 3 & & & & & & & & & & & & & \multicolumn{1}{|r}{0.0} & 0.0 & 100.0 &  \multicolumn{1}{r|}{0.0}  \\
                          & Full $\rightarrow$ Assoc  & 4 &  & & &  &  & & & & & & & & \multicolumn{1}{|r}{0.0} & 25.0 & 75.0 &  \multicolumn{1}{r|}{0.0}  \\
                          & Full $\rightarrow$ Full   & 158 &  &  &  &  &  &  &  &  &  &  &  &  & \multicolumn{1}{|r}{0.0} & 0.0 & 98.7 &  \multicolumn{1}{r|}{1.3}  \\	
                          & Full $\rightarrow$ Gone   & 65 &  &  &  &  &  &  &  &  &  &  &  &  & \multicolumn{1}{|r}{0.0} & 0.0 & 69.2 &  \multicolumn{1}{r|}{30.8} \\ \cline{2-19}
                          & Total & 607 &  &  &  &  &  &  &  &  &  &  &  &  &  &  &  &               
\end{tabular}
\caption{{\bf Estimated error rates for faculty rank transitions from 2011 to 2017.} {\normalfont Estimated error rates (expressed as percentages; rows sum to 100\%) for all possible transitions of the form $X \to Y$, where $X$ is the rank of a faculty member in the 2011 manual census (where ``New'' indicates that they were unobserved in 2011) and $Y$ is their rank in 2017 (where ``Gone'' indicates that they were unobserved at any institution in 2017). To construct this confusion matrix, we used crowdsourcing to determine $Y$ for a 10\% random sample of the 2011 faculty, and compared those titles (columns) to the output of our crawler (rows).}}
%\caption{Estimated error rates (expressed as percentages) for all possible transitions of the form $X \to Y$, where $X$ is the rank of a faculty member in the 2011 manual census (where ``New'' indicates that they were unobserved in 2011) and $Y$ is their rank in the 2017 automated census (where ``Gone'' indicates that they were unobserved in 2017). This confusion matrix was tabulated using a majority rule aggregation for labels given by crowd workers asked to identify the 2017 rank of a uniformly random 10\% sample of 2011 faculty of different ranks, and was used to correct the estimates in the 2017 census.} 
\label{tab:confusion}
\end{table*}
%\end{comment}
% --------------------

\subsection{Deploying and evaluating the crawler}
\label{sec:deploying}

We now evaluate the performance of the entire system, applied to the full set of 205 computer science departments. Hence, the system now starts from each department homepage, navigates to its directory, parses all pages it classifies as being a potential directory, and finally writes out a directory of all TTT faculty. Running as a single-threaded process on an off-the-shelf laptop, the overall time required to produce structured directories for all 205 computer science departments is roughly 3 CPU hours. The majority of this time is spent parsing directories, which could be potentially reduced using a more accurate directory classifier.

Compared to the 2011 manually collected census~\cite{clauset:hiring}, which took 1600 hours of work by a team of 13 data collectors, our automated approach %, which runs to completion in a little over 3 CPU hours, 
is substantially more efficient. In fact, the average time required to produce a single department's faculty directory is 55 CPU seconds. Launching 205 instances of our crawler, one for each department, in a modern cloud-computing environment should lower the running time to under a minute total. In such a setting, a full census of an academic field can be automatically assembled nearly 100,000 times faster than by hand.

However, for 509 professors (10\%) of the 2017 census, our system could not obtain an unambiguous title from the departmental faculty listings. For instance, some directories include lists or tables of the names of faculty members with nothing more specific than headings like ``full-time", ``tenure-track" or ``professors". We obtain these missing titles using crowd workers on Mechanical Turk. In a production-like environment, an automated system like ours would likely need to be complemented by a small amount of human labeling to correct such errors and missing information.

Our 2017 census of North American computer science departments contains 4838 faculty members: 2474 (51.1\%) full professors, 1298 (26.8\%) associate professors, and 1066 (22.0\%) assistant professors.

% ----- FIGURE -----
\begin{figure}[b!]
\includegraphics[width=\columnwidth]{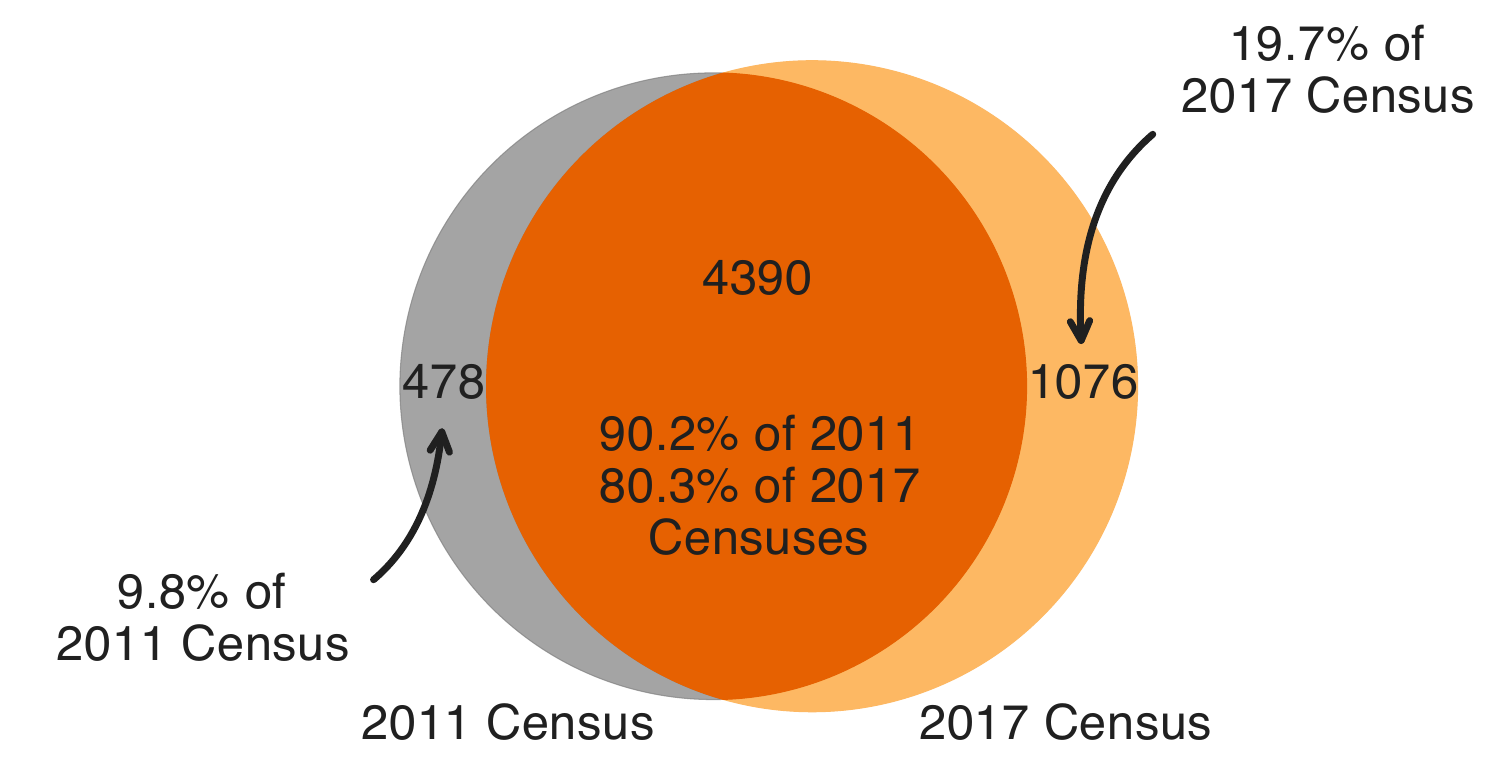}
\caption{{\bf Faculty overlap in 2011 and 2017 censuses, adjusted for errors.} {\normalfont The automatically collected 2017 census includes over 90\% of the faculty from the manually curated 2011 census. Non-overlapping faculty counts align with reported growth estimate from the CRA (see main text).}}
%\caption{Overlap of faculty between 2011 and 2017 censuses, adjusted for errors. Our automatically collected census includes over 90\% of the 2011 manually curated census.}
\label{fig:overlap}
\end{figure}
% --------------------

\subsection{Extending to other fields}
\label{sec:extending}

%More generally, access to the kind of census data produced by our crawler should enable many new and exciting research directions while also enhancing existing efforts. Often work studying academics focuses on researchers of a single academic field~\cite{cole:citationprestige}; applying our system to multiple fields would allow researchers to investigate relationships across disciplines, e.g., how computing faculty are moving into other domains. 
To test the generality of our system on other academic fields, we have made a preliminary application of our system to 144 history departments and 112 business schools, both of which were also part of the 2011 manual census~\cite{clauset:hiring}. Our results suggest that relatively little customization is needed to adapt the system to other academic fields. Specifically, we visited the online directories for each of these academic units, selected the first person listed, and checked whether our 2017 automated census of these fields contained a record for that person. In 82\% of history departments and 77\% of business schools' directories, we correctly recalled these faculty members with no modifications to the system. Errors here were caused by particularly complicated (often multi-page, separated by sub-disciplines) directory formats or novel faculty names, both easily corrected, and not by novel faculty titles. The loss of accuracy due to faculty names is easily addressed by incorporating a more exhaustive list of names, e.g., all surnames recorded by the U.S. Census. Parsing novel directory structures will require modest additional software development to recognize and navigate these other forms of HTML pagination. Multi-page directories are uncommon in computer science, but more common in larger fields like business schools, and it should be straightforward to extend our crawler system to handle these more complicated formats.

\section{Retention in computer science}
\label{sec:science}

Having applied our system to the same 205 PhD-granting computer science departments as the 2011 manually-collected census~\cite{clauset:hiring}, we can now compare this 6-year old snapshot with our 2017 automatically-generated census. This comparison illustrates the utility of a system for automatically assembling an academic census and allows us to characterize the kinds of errors it makes. We also use this comparison to quantify recent turnover and retention of computing faculty. We first perform this analysis for faculty as a whole, and then consider turnover and retention for female faculty specifically. This latter step allows us to provide new insights into a question of broad relevance in computing: Are women leaving the professoriate at greater rates than men? 

In order to make our comparison fair and to improve the accuracy of our estimates of turnover and retention rates, a few additional post-processing steps were necessary. Of the 205 departments surveyed, 16 are Departments of Electrical Engineering and Computer Science (EECS) and 30 are Departments of Computer Science and Engineering (CSE), meaning that their faculty directories included both CS and EE faculty. The 2011 census manually separated and removed the EE faculty, and we repeat this process on the results of our system for the same departments, using faculty research interests as the separating variables. We then performed a simple matching based on the first initial and last name strings of 2011 faculty and 2017 faculty. The results of this operation divided the set of all faculty into three groups: (i) new faculty (1329 absent in 2011 and present in 2017), (ii) retained faculty (3509 present in 2011 and in 2017), and (iii) departed faculty (1248 present in 2011 and absent in 2017). We validated this matching procedure and the accuracy of the identified ranks of faculty by using crowdsourcing to obtain the current positions for uniformly random 10\% samples of each of the assistant, associate, and full professor groups from the 2011 census (475 faculty total). Each current position was collected twice and 108 observed disagreements were then manually evaluated, producing a majority vote label aggregation. Additionally, we manually checked a uniformly random 10\% (132) of the 2017 assistant, associate, and full professors who were new %to our data set 
(not seen in 2011). The results of these efforts were tabulated in a confusion matrix representing the error rates for classifying faculty by their faculty rank and by their membership in the new, retained, and departed groups (see Table~\ref{tab:confusion}).

This confusion matrix was then used to derive corrected counts for faculty by rank and membership, multiplying the distribution of transitions generated from our crawler by the MTurker's estimated transition rates. Aggregating these corrected counts across ranks yields
%transitions from status $X$ in 2011 to status $Y$ in 2017, where a status could be ``new,'' Assistant, Associate, or Full Professor, or ``departed''
%all possible transitions, and was used to derive corrected counts for faculty in the three groups: 
1076 new hires, 4390 retained faculty, and 478 departed (not observed at any in-sample institution) faculty. Overall, we find that 90.2\% of faculty observed in 2011 are also found in our 2017 census (Fig.~\ref{fig:overlap}). Furthermore, the number of new hires (19.7\%) is more than twice as large as the number of departed faculty (9.8\%), reflecting the overall growth in computing over this time period.

The CRA provides estimates of both department growth and losses based on information provided by a survey of  the heads of departments.  According to the CRA's 2011 and 2017 estimates of the number of employed tenure-track faculty from all US and Canadian CS departments,\footnote{Table F1: {\scriptsize \url{cra.org/resources/taulbee-survey/}} } there was an 11\% growth in number of faculty. We find comparable net growth (12\%) in the total number of computing faculty over 6 years. From 2012--2016, the CRA reports a total of 1206 computing faculty who left their existing positions, with 818 of these leaving academia entirely.\footnote{Table F5: {\scriptsize \url{cra.org/resources/taulbee-survey/}} } The size of the departed group is quite small compared to the CRA's own estimate of total faculty losses. This discrepancy likely stems from the fact that the CRA's data come from a social survey, while ours come directly from online directories and web searches. For instance, the CRA does not capture information about faculty who leave and then return to academia, while these faculty would appear in our data. A useful line of future work would involve a deeper comparison of the CRA's surveys with our faculty directory information.
%From 2012--2016, the CRA reports a total of 696 computing faculty who left their existing positions because they took a non-academic position, transitioned to a part-time role, retired or died.

% ----- FIGURE -----
\begin{figure*}[t!]
\includegraphics[width=\textwidth]{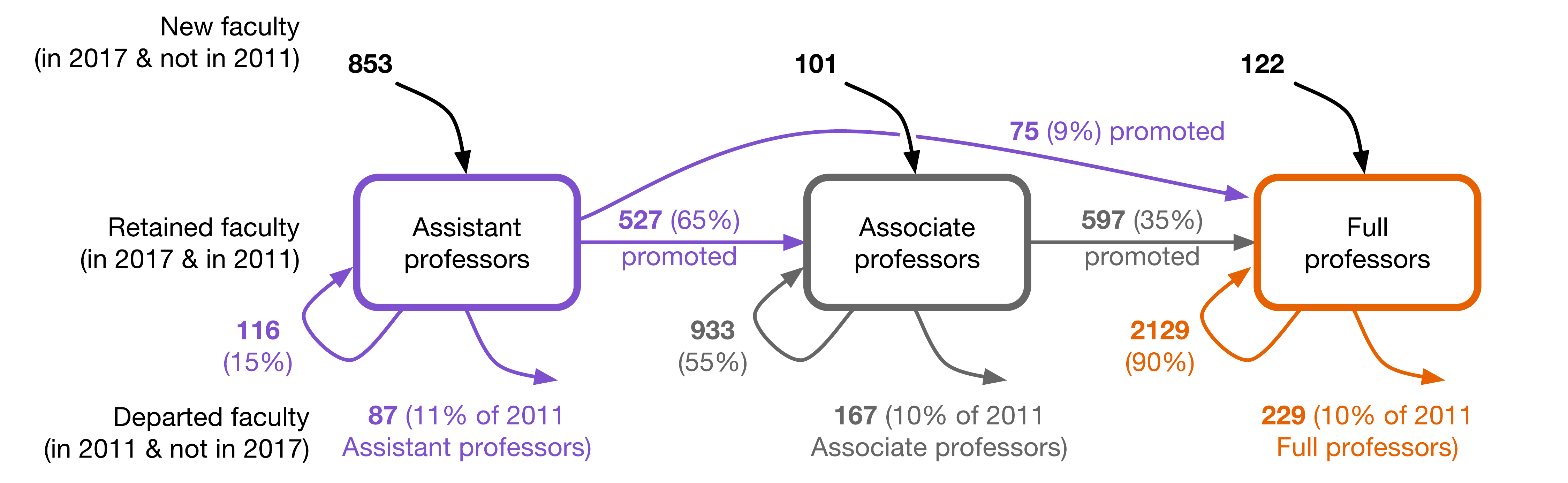}
\caption{{\bf Faculty title transitions between 2011 and 2017 censuses.} {\normalfont Flows of computer science faculty into, among, and out of the assistant, associate, and full professor ranks, comparing the 2011 manual census with the 2017 automated census. Counts are corrected for sampling errors in 2017 (see main text). Flows representing less than 1\% of all faculty are omitted for clarity.}}
%\caption{Flows of computer science faculty into, among, and out of the assistant, associate, and full professor ranks, derived by comparing the 2011 manual census with the 2017 automated census. Counts are corrected for sampling errors in 2017. Flows representing less than 1\% of all faculty are omitted for clarity.}
\label{fig:pipeline}
\end{figure*}
% --------------------

Subdividing our three faculty groups (new hires, retained faculty, and departed faculty) according to each faculty member's rank (assistant, associate, or full) in 2011 and 2017, we can examine the flows of faculty into, through, and out of different career stages (Fig.~\ref{fig:pipeline}). %Again, we match on first initial and last name and adjust our values based on the error rates which we collected above. 
Reflecting our finding of a substantial net growth in faculty, there is relatively large inflow of new assistant professors, and large retention of associate and full professors. It is notable that the outflow rate of assistant professors is comparable to the outflow rates of associate and full professors. Na\"ively, we might have expected the outflow to be larger at the assistant professor stage, reflecting the impact of negative tenure decisions.

% ----- TABLE -----
\begin{table}[b!]
\begin{tabular}{cccc|c|cccc}
                                  &                                      &&             &          & \multicolumn{4}{c}{2017}                \\
                                  &                                      &&             & $n$       & Asst    & Assoc  & Full      & Gone  \\ \cline{4-9}
\multirow{8}{*}{\rotatebox{90}{2011}} &\multirow{3}{*}{\rotatebox{90}{Women}} && Asst     & 254   & 0.299  & 0.480   & 0.147  & 0.074  \\
                                  &                                      && Assoc  & 174   & 0.180  & 0.437   & 0.096 & 0.287  \\
                                  &                                      && Full      & 263   & 0.000  & 0.005  & 0.898  & 0.098  \\ \cline{4-9}  
                                  %&                                      &&             &          &            &             &            & 0.137 \\ 
                                  & & & & & \\ \cline{4-9}
                                  &\multirow{3}{*}{\rotatebox{90}{Men}}      && Asst     & 1408 & 0.308  & 0.470   & 0.129  & 0.094  \\
                                  &                                      && Assoc  & 618   & 0.162  & 0.433   & 0.131  & 0.274  \\
                                  &                                      && Full      & 2034 & 0.000  & 0.002  & 0.898  & 0.010 \\ \cline{4-9}
                                  %&                                      &&             &          &            &             &            & 0.124 \\
\end{tabular}
\vspace{2mm}
\caption{{\bf Faculty title transition probabilities differ slightly between men and women.} {\normalfont Transition matrix showing the probability, based on corrected counts, that a female or male faculty member has one rank in 2011 and another in 2017. ``Gone'' indicates faculty not observed at any university in 2017, and this column gives the rank-level attrition rates of 2011 faculty. Total attrition rates are 0.137 for women and 0.124 for men.}}
%\caption{Transition matrix showing the probability, based on corrected counts, that a female or male faculty member has one rank in 2011 and another rank in 2017. ``Gone'' indicates the person was unobserved in 2017, and this column gives the rank-level attrition rates of 2011 faculty. The total attrition rates are 0.137 for women and 0.124 for men.} 
\label{tab:transitions}
\end{table}
% --------------------

Finally, the 2011 manual census also includes information about each professor's gender, allowing us to estimate gender differences in rates of retention, promotion, and attrition within the CS tenure-track pipeline (Table~\ref{tab:transitions}). These counts indicate that slightly more women than men were retained from the group of 2011 assistant professors (92.6\% vs.\ 90.6\%). At the same time, fewer women than men were retained from the groups of 2011 associate professors (71.3\% vs. 72.6\%) and full professors (90.2\% vs. 99.0\%). Future work should involve investigating the gender differences among new faculty, since the 2017 census does not contain information about gender.

Aggregating across ranks, attrition rates for women and men are similar (13.7\% vs.\ 12.4\%), but slightly higher for women. This modest difference is consistent with the ``leaky pipeline,'' a metaphor stemming from a large body of literature showing that women leave academia at slightly higher rates than men at all stages of an academic career~\cite{xu2008gender,kulis2002more,pell1996fixing}, including computer science in particular~\cite{jadidi2017gender}. A key question, however, is whether these observed differences can be attributed to fluctuations. Our data cannot definitively answer this question. However, if we model the rates of retention and promotions across gender as independent random variables, then under a binomial test for each transition, the women's attrition rate is not significantly different from the men's ($p=0.36$). (We also do not find any significant difference under a $\chi^2$ test, $p=0.40$.)
%That is, these differences in attrition rates are not statistically significant. 
That said, a standard hypothesis test may make unrealistic assumptions about independence in this setting, and so the lack of significance in comparing two somewhat arbitrarily dated snapshots should not be over interpreted. Longitudinal analysis of, for example, yearly censuses is surely necessary in order to correctly evaluate the true significance of the observed differences. An automated system like the one presented here should make that possible moving forward.

\section{Conclusion}
\label{sec:conclusion}

The ability to cheaply and quickly assemble a complete census of an academic field from web-based data will accelerate research on a wide variety of social and policy questions about the composition, dynamics, and diversity of the scientific workforce in general, and computing fields in particular. The past difficulty of performing such a census %in the past 
has limited such %research 
efforts, %in these directions, 
and researchers have instead used less reliable survey or sampling methods. The novel system we describe here, which uses a topical web crawler to automatically assemble an academic census from semi-structured web-based data, is both accurate and efficient. In a modern cloud computing environment, this system could essentially run at scale in realtime, on as many fields as desired.

The modular design of our system enables independent incremental improvements to its overall performance, e.g., by developing better techniques for parsing the semi-structured information stored in departmental faculty listings or for selecting target individuals out of the full listings. That said, the high precision and recall of the system when applied to North American PhD-granting computer science departments suggests that it is already quite effective. We now focus on the limitations of our current system and outline specific recommendations for how future studies might enhance and extend our work to other disciplines.

First, the system's specification currently requires several hand-constructed whitelists or blacklists, or manual interventions in order to achieve high accuracy. An important direction for future work would be to automate these steps. For example, identifying which faculty members in a departmental listing are in-sample for a particular academic field can require manual investigation, as in the case of distinguishing EE versus CS faculty in our study. Any application to the biological or biomedical sciences would also require such separation, as the corresponding disciplines are mixed in complicated ways across many departments. This step could be automated to some degree by using topic models to cluster faculty interests based on their publications~\cite{way:gender} or on their collaboration or citation networks~\cite{rosvall:map:2008}. Automating the discovery of distinguishing features would also drive the system's expansion to other languages, enabling new studies of the increasingly-international scientific workforce.

Our system was unable to identify the faculty rank for about 10\% of in-sample faculty, and we collected this information manually via crowd work. An easy way to improve the system's performance in this direction would be to perform deeper crawling for each identified faculty member, e.g., crawling their professional homepage, parsing their curriculum vitae, or performing targeted web queries. The information gained through this additional work would need to be evaluated carefully, however, as different sources of information will have different levels of authority or recency.

For a system like this, some amount of manual work is essential in order to detect, characterize, and correct the census's errors. The detailed evaluation we performed in our comparison of our automated 2017 census of computer science with the manual 2011 census illustrates this point well, as the confusion matrix we constructed via crowd work allowed us to obtain more accurate estimates of counts of faculty at different ranks in 2017. Ideally, a more accurate automated system would make fewer such human-measurable errors, and constructing such a matrix serves to highlight where accuracy improvements could be made.

The large overlap between our system's 2017 census and the manual 2011 census demonstrates the utility of a cheap and efficient automated census system. We find close agreement between the CRA's official survey-based estimate of the net growth of computing faculty and our own automated estimate. Our analysis of the flows into, out of, and through faculty ranks overall, and for female faculty in particular, demonstrates that an automated census can provide detailed insights on important questions about the composition and dynamics of the scientific workforce. A thorough investigation of the patterns we observe, including the observation that slightly more female than male assistant professors from 2011 were retained as of 2017, while substantially fewer female full professors were retained (Table 2), would require a longitudinal study. Such a multi-year census effort should now be straightforward using the system described here.

As was evident in our analysis of the retention of female faculty from the 2011 census of computer science, a key future direction will be the development of longitudinal data, which would allow more detailed investigations of trends in hiring, promotion, retention, and attrition. The system presented here is fast and suitable for continuous collection of faculty employment information over time. It could also be adapted to historical faculty listings stored in the Internet Archive.%
\footnote{See {\scriptsize \url{archive.org/web/}} }
We look forward to these and other developments, and the many scientific insights that will come from having an inexpensive and accurate method for automatically assembling a full census of an academic field.\\

\begin{acknowledgements}\vspace{-2mm}
The authors thank Mirta Galesic and Daniel Larremore for helpful conversations. All authors were supported by NSF award SMA~1633747.
\end{acknowledgements}


\begin{thebibliography}{37}%
\makeatletter
\providecommand \@ifxundefined [1]{%
 \@ifx{#1\undefined}
}%
\providecommand \@ifnum [1]{%
 \ifnum #1\expandafter \@firstoftwo
 \else \expandafter \@secondoftwo
 \fi
}%
\providecommand \@ifx [1]{%
 \ifx #1\expandafter \@firstoftwo
 \else \expandafter \@secondoftwo
 \fi
}%
\providecommand \natexlab [1]{#1}%
\providecommand \enquote  [1]{``#1''}%
\providecommand \bibnamefont  [1]{#1}%
\providecommand \bibfnamefont [1]{#1}%
\providecommand \citenamefont [1]{#1}%
\providecommand \href@noop [0]{\@secondoftwo}%
\providecommand \href [0]{\begingroup \@sanitize@url \@href}%
\providecommand \@href[1]{\@@startlink{#1}\@@href}%
\providecommand \@@href[1]{\endgroup#1\@@endlink}%
\providecommand \@sanitize@url [0]{\catcode `\\12\catcode `\$12\catcode
  `\&12\catcode `\#12\catcode `\^12\catcode `\_12\catcode `\%12\relax}%
\providecommand \@@startlink[1]{}%
\providecommand \@@endlink[0]{}%
\providecommand \url  [0]{\begingroup\@sanitize@url \@url }%
\providecommand \@url [1]{\endgroup\@href {#1}{\urlprefix }}%
\providecommand \urlprefix  [0]{URL }%
\providecommand \Eprint [0]{\href }%
\providecommand \doibase [0]{http://dx.doi.org/}%
\providecommand \selectlanguage [0]{\@gobble}%
\providecommand \bibinfo  [0]{\@secondoftwo}%
\providecommand \bibfield  [0]{\@secondoftwo}%
\providecommand \translation [1]{[#1]}%
\providecommand \BibitemOpen [0]{}%
\providecommand \bibitemStop [0]{}%
\providecommand \bibitemNoStop [0]{.\EOS\space}%
\providecommand \EOS [0]{\spacefactor3000\relax}%
\providecommand \BibitemShut  [1]{\csname bibitem#1\endcsname}%
\let\auto@bib@innerbib\@empty
%</preamble>
\bibitem [{\citenamefont {Van~Dyne}\ and\ \citenamefont
  {Saavedra}(1996)}]{vandyne1996conflict}%
  \BibitemOpen
  \bibfield  {author} {\bibinfo {author} {\bibfnamefont {Linn}\ \bibnamefont
  {Van~Dyne}}\ and\ \bibinfo {author} {\bibfnamefont {Richard}\ \bibnamefont
  {Saavedra}},\ }\bibfield  {title} {\enquote {\bibinfo {title} {A naturalistic
  minority influence experiment: Effects on divergent thinking, conflict and
  originality in work-groups},}\ }\href@noop {} {\bibfield  {journal} {\bibinfo
   {journal} {Br.\ J.\ Soc.\ Psychol}\ }\textbf {\bibinfo {volume} {35}},\
  \bibinfo {pages} {151--167} (\bibinfo {year} {1996})}\BibitemShut {NoStop}%
\bibitem [{\citenamefont {McLeod}\ \emph {et~al.}(1996)\citenamefont {McLeod},
  \citenamefont {Lobel},\ and\ \citenamefont {Cox~Jr}}]{mcleod1996ethnic}%
  \BibitemOpen
  \bibfield  {author} {\bibinfo {author} {\bibfnamefont {Poppy~Lauretta}\
  \bibnamefont {McLeod}}, \bibinfo {author} {\bibfnamefont {Sharon~Alisa}\
  \bibnamefont {Lobel}}, \ and\ \bibinfo {author} {\bibfnamefont {Taylor~H}\
  \bibnamefont {Cox~Jr}},\ }\bibfield  {title} {\enquote {\bibinfo {title}
  {Ethnic diversity and creativity in small groups},}\ }\href@noop {}
  {\bibfield  {journal} {\bibinfo  {journal} {Small Group Res.}\ }\textbf
  {\bibinfo {volume} {27}},\ \bibinfo {pages} {248--264} (\bibinfo {year}
  {1996})}\BibitemShut {NoStop}%
\bibitem [{\citenamefont {Page}(2008)}]{page2008}%
  \BibitemOpen
  \bibfield  {author} {\bibinfo {author} {\bibfnamefont {Scott~E}\ \bibnamefont
  {Page}},\ }\href@noop {} {\emph {\bibinfo {title} {The Difference: How the
  Power of Diversity Creates Better Groups, Firms, Schools, and Societies}}}\
  (\bibinfo  {publisher} {Princeton University Press},\ \bibinfo {year}
  {2008})\BibitemShut {NoStop}%
\bibitem [{\citenamefont {Milem}(2003)}]{milem2003educational}%
  \BibitemOpen
  \bibfield  {author} {\bibinfo {author} {\bibfnamefont {Jeffrey~F}\
  \bibnamefont {Milem}},\ }\bibfield  {title} {\enquote {\bibinfo {title} {The
  educational benefits of diversity: Evidence from multiple sectors},}\ }in\
  \href@noop {} {\emph {\bibinfo {booktitle} {Compelling Interest:\ Examining
  the evidence on racial dynamics in higher education}}},\ \bibinfo {editor}
  {edited by\ \bibinfo {editor} {\bibfnamefont {Mitchel~J}\ \bibnamefont
  {Chang}}, \bibinfo {editor} {\bibfnamefont {Daria}\ \bibnamefont {Witt}},
  \bibinfo {editor} {\bibfnamefont {James}\ \bibnamefont {Jones}}, \ and\
  \bibinfo {editor} {\bibfnamefont {Kenji}\ \bibnamefont {Hakuta}}}\ (\bibinfo
  {publisher} {Stanford University Press},\ \bibinfo {year} {2003})\ pp.\
  \bibinfo {pages} {126--169}\BibitemShut {NoStop}%
\bibitem [{\citenamefont {Hill}\ \emph {et~al.}(2010)\citenamefont {Hill},
  \citenamefont {Corbett},\ and\ \citenamefont {St~Rose}}]{hill2010so}%
  \BibitemOpen
  \bibfield  {author} {\bibinfo {author} {\bibfnamefont {Catherine}\
  \bibnamefont {Hill}}, \bibinfo {author} {\bibfnamefont {Christianne}\
  \bibnamefont {Corbett}}, \ and\ \bibinfo {author} {\bibfnamefont {Andresse}\
  \bibnamefont {St~Rose}},\ }\href@noop {} {\emph {\bibinfo {title} {Why so
  few? Women in science, technology, engineering, and mathematics}}}\ (\bibinfo
   {publisher} {American Association of University Women},\ \bibinfo {year}
  {2010})\BibitemShut {NoStop}%
\bibitem [{\citenamefont {Way}\ \emph {et~al.}(2016)\citenamefont {Way},
  \citenamefont {Larremore},\ and\ \citenamefont {Clauset}}]{way:gender}%
  \BibitemOpen
  \bibfield  {author} {\bibinfo {author} {\bibfnamefont {Samuel~F}\
  \bibnamefont {Way}}, \bibinfo {author} {\bibfnamefont {Daniel~B}\
  \bibnamefont {Larremore}}, \ and\ \bibinfo {author} {\bibfnamefont {Aaron}\
  \bibnamefont {Clauset}},\ }\bibfield  {title} {\enquote {\bibinfo {title}
  {Gender, productivity, and prestige in computer science faculty hiring
  networks},}\ }in\ \href@noop {} {\emph {\bibinfo {booktitle} {Proc.\ 25th
  Internat.\ Conf.\ on World Wide Web (WWW)}}}\ (\bibinfo {year} {2016})\ pp.\
  \bibinfo {pages} {1169--1179}\BibitemShut {NoStop}%
\bibitem [{\citenamefont {Fowler}\ \emph {et~al.}(2007)\citenamefont {Fowler},
  \citenamefont {Grofman},\ and\ \citenamefont {Masuoka}}]{fowler2007social}%
  \BibitemOpen
  \bibfield  {author} {\bibinfo {author} {\bibfnamefont {James~H}\ \bibnamefont
  {Fowler}}, \bibinfo {author} {\bibfnamefont {Bernard}\ \bibnamefont
  {Grofman}}, \ and\ \bibinfo {author} {\bibfnamefont {Natalie}\ \bibnamefont
  {Masuoka}},\ }\bibfield  {title} {\enquote {\bibinfo {title} {Social networks
  in political science:\ hiring and placement of {PhDs}, 1960--2002},}\
  }\href@noop {} {\bibfield  {journal} {\bibinfo  {journal} {PS: Political
  Science \& Politics}\ }\textbf {\bibinfo {volume} {40}},\ \bibinfo {pages}
  {729--739} (\bibinfo {year} {2007})}\BibitemShut {NoStop}%
\bibitem [{\citenamefont {Cole}(1979)}]{cole1979age}%
  \BibitemOpen
  \bibfield  {author} {\bibinfo {author} {\bibfnamefont {Stephen}\ \bibnamefont
  {Cole}},\ }\bibfield  {title} {\enquote {\bibinfo {title} {Age and scientific
  performance},}\ }\href@noop {} {\bibfield  {journal} {\bibinfo  {journal}
  {Am.\ J.\ Sociol.}\ ,\ \bibinfo {pages} {958--977}} (\bibinfo {year}
  {1979})}\BibitemShut {NoStop}%
\bibitem [{\citenamefont {Allison}\ and\ \citenamefont
  {Long}(1990)}]{allison:departmenteffects}%
  \BibitemOpen
  \bibfield  {author} {\bibinfo {author} {\bibfnamefont {Paul~D}\ \bibnamefont
  {Allison}}\ and\ \bibinfo {author} {\bibfnamefont {J~Scott}\ \bibnamefont
  {Long}},\ }\bibfield  {title} {\enquote {\bibinfo {title} {Departmental
  effects on scientific productivity},}\ }\href@noop {} {\bibfield  {journal}
  {\bibinfo  {journal} {Am.\ Sociol.\ Rev}\ }\textbf {\bibinfo {volume} {55}},\
  \bibinfo {pages} {469--478} (\bibinfo {year} {1990})}\BibitemShut {NoStop}%
\bibitem [{\citenamefont {Long}\ and\ \citenamefont
  {Fox}(1995)}]{long1995scientific}%
  \BibitemOpen
  \bibfield  {author} {\bibinfo {author} {\bibfnamefont {J~Scott}\ \bibnamefont
  {Long}}\ and\ \bibinfo {author} {\bibfnamefont {Mary~Frank}\ \bibnamefont
  {Fox}},\ }\bibfield  {title} {\enquote {\bibinfo {title} {Scientific careers:
  Universalism and particularism},}\ }\href@noop {} {\bibfield  {journal}
  {\bibinfo  {journal} {Annu.\ Rev.\ Sociol.}\ }\textbf {\bibinfo {volume}
  {21}},\ \bibinfo {pages} {45--71} (\bibinfo {year} {1995})}\BibitemShut
  {NoStop}%
\bibitem [{\citenamefont {Xie}\ and\ \citenamefont {Shauman}(1998)}]{xie98}%
  \BibitemOpen
  \bibfield  {author} {\bibinfo {author} {\bibfnamefont {Yu}~\bibnamefont
  {Xie}}\ and\ \bibinfo {author} {\bibfnamefont {Kimberlee~A}\ \bibnamefont
  {Shauman}},\ }\bibfield  {title} {\enquote {\bibinfo {title} {Sex differences
  in research productivity: New evidence about an old puzzle},}\ }\href@noop {}
  {\bibfield  {journal} {\bibinfo  {journal} {Am.\ Sociol.\ Rev.}\ }\textbf
  {\bibinfo {volume} {63}},\ \bibinfo {pages} {847--870} (\bibinfo {year}
  {1998})}\BibitemShut {NoStop}%
\bibitem [{\citenamefont {Myers}\ \emph {et~al.}(2011)\citenamefont {Myers},
  \citenamefont {Mucha},\ and\ \citenamefont {Porter}}]{myers2011mathematical}%
  \BibitemOpen
  \bibfield  {author} {\bibinfo {author} {\bibfnamefont {Sean~A}\ \bibnamefont
  {Myers}}, \bibinfo {author} {\bibfnamefont {Peter~J}\ \bibnamefont {Mucha}},
  \ and\ \bibinfo {author} {\bibfnamefont {Mason~A}\ \bibnamefont {Porter}},\
  }\bibfield  {title} {\enquote {\bibinfo {title} {Mathematical genealogy and
  department prestige},}\ }\href@noop {} {\bibfield  {journal} {\bibinfo
  {journal} {Chaos}\ }\textbf {\bibinfo {volume} {21}},\ \bibinfo {pages}
  {041104} (\bibinfo {year} {2011})}\BibitemShut {NoStop}%
\bibitem [{\citenamefont {Amir}\ and\ \citenamefont
  {Knauff}(2008)}]{amir2008ranking}%
  \BibitemOpen
  \bibfield  {author} {\bibinfo {author} {\bibfnamefont {Rabah}\ \bibnamefont
  {Amir}}\ and\ \bibinfo {author} {\bibfnamefont {Malgorzata}\ \bibnamefont
  {Knauff}},\ }\bibfield  {title} {\enquote {\bibinfo {title} {Ranking
  economics departments worldwide on the basis of {PhD} placement},}\
  }\href@noop {} {\bibfield  {journal} {\bibinfo  {journal} {Rev.\ Econ.\
  Stat}\ }\textbf {\bibinfo {volume} {90}},\ \bibinfo {pages} {185--190}
  (\bibinfo {year} {2008})}\BibitemShut {NoStop}%
\bibitem [{\citenamefont {Katz}\ \emph {et~al.}(2011)\citenamefont {Katz},
  \citenamefont {Gubler}, \citenamefont {Zelner}, \citenamefont {Bommarito},
  \citenamefont {Provins},\ and\ \citenamefont
  {Ingall}}]{katz2011reproduction}%
  \BibitemOpen
  \bibfield  {author} {\bibinfo {author} {\bibfnamefont {Daniel~Martin}\
  \bibnamefont {Katz}}, \bibinfo {author} {\bibfnamefont {Joshua~R}\
  \bibnamefont {Gubler}}, \bibinfo {author} {\bibfnamefont {Jon}\ \bibnamefont
  {Zelner}}, \bibinfo {author} {\bibfnamefont {Michael~J}\ \bibnamefont
  {Bommarito}}, \bibinfo {author} {\bibfnamefont {Eric}\ \bibnamefont
  {Provins}}, \ and\ \bibinfo {author} {\bibfnamefont {Eitan}\ \bibnamefont
  {Ingall}},\ }\bibfield  {title} {\enquote {\bibinfo {title} {Reproduction of
  hierarchy? a social network analysis of the {American} law professoriate},}\
  }\href@noop {} {\bibfield  {journal} {\bibinfo  {journal} {J.\ Legal Educ}\
  }\textbf {\bibinfo {volume} {61}},\ \bibinfo {pages} {76--103} (\bibinfo
  {year} {2011})}\BibitemShut {NoStop}%
\bibitem [{\citenamefont {Schmidt}\ and\ \citenamefont
  {Chingos}(2007)}]{schmidt2007ranking}%
  \BibitemOpen
  \bibfield  {author} {\bibinfo {author} {\bibfnamefont {Benjamin~M}\
  \bibnamefont {Schmidt}}\ and\ \bibinfo {author} {\bibfnamefont {Matthew~M}\
  \bibnamefont {Chingos}},\ }\bibfield  {title} {\enquote {\bibinfo {title}
  {Ranking doctoral programs by placement:\ a new method},}\ }\href@noop {}
  {\bibfield  {journal} {\bibinfo  {journal} {PS: Political Science \&
  Politics}\ }\textbf {\bibinfo {volume} {40}},\ \bibinfo {pages} {523--529}
  (\bibinfo {year} {2007})}\BibitemShut {NoStop}%
\bibitem [{\citenamefont {Hanneman}(2001)}]{hanneman2001prestige}%
  \BibitemOpen
  \bibfield  {author} {\bibinfo {author} {\bibfnamefont {Robert~A}\
  \bibnamefont {Hanneman}},\ }\bibfield  {title} {\enquote {\bibinfo {title}
  {The prestige of {Ph.D.} granting departments of sociology: A simple network
  approach},}\ }\href@noop {} {\bibfield  {journal} {\bibinfo  {journal}
  {Connections}\ }\textbf {\bibinfo {volume} {24}},\ \bibinfo {pages} {68--77}
  (\bibinfo {year} {2001})}\BibitemShut {NoStop}%
\bibitem [{\citenamefont {Zuckerman}(1977)}]{zuckerman1977scientific}%
  \BibitemOpen
  \bibfield  {author} {\bibinfo {author} {\bibfnamefont {Harriet}\ \bibnamefont
  {Zuckerman}},\ }\href@noop {} {\emph {\bibinfo {title} {Scientific elite:
  Nobel laureates in the United States}}}\ (\bibinfo  {publisher} {Transaction
  Publishers},\ \bibinfo {year} {1977})\BibitemShut {NoStop}%
\bibitem [{\citenamefont {Schlagberger}\ \emph {et~al.}(2016)\citenamefont
  {Schlagberger}, \citenamefont {Bornmann},\ and\ \citenamefont
  {Bauer}}]{schlagberger2016institutions}%
  \BibitemOpen
  \bibfield  {author} {\bibinfo {author} {\bibfnamefont {Elisabeth~Maria}\
  \bibnamefont {Schlagberger}}, \bibinfo {author} {\bibfnamefont {Lutz}\
  \bibnamefont {Bornmann}}, \ and\ \bibinfo {author} {\bibfnamefont {Johann}\
  \bibnamefont {Bauer}},\ }\bibfield  {title} {\enquote {\bibinfo {title} {At
  what institutions did {Nobel} laureates do their prize-winning work? {An}
  analysis of biographical information on {Nobel} laureates from 1994 to
  2014},}\ }\href@noop {} {\bibfield  {journal} {\bibinfo  {journal}
  {Scientometrics}\ }\textbf {\bibinfo {volume} {109}},\ \bibinfo {pages}
  {723--767} (\bibinfo {year} {2016})}\BibitemShut {NoStop}%
\bibitem [{\citenamefont {Clauset}\ \emph {et~al.}(2015)\citenamefont
  {Clauset}, \citenamefont {Arbesman},\ and\ \citenamefont
  {Larremore}}]{clauset:hiring}%
  \BibitemOpen
  \bibfield  {author} {\bibinfo {author} {\bibfnamefont {Aaron}\ \bibnamefont
  {Clauset}}, \bibinfo {author} {\bibfnamefont {Samuel}\ \bibnamefont
  {Arbesman}}, \ and\ \bibinfo {author} {\bibfnamefont {Daniel~B}\ \bibnamefont
  {Larremore}},\ }\bibfield  {title} {\enquote {\bibinfo {title} {{Systematic
  inequality and hierarchy in faculty hiring networks}},}\ }\href@noop {}
  {\bibfield  {journal} {\bibinfo  {journal} {Sci.\ Adv}\ }\textbf {\bibinfo
  {volume} {1}},\ \bibinfo {pages} {e1400005} (\bibinfo {year}
  {2015})}\BibitemShut {NoStop}%
\bibitem [{\citenamefont {Way}\ \emph {et~al.}(2017)\citenamefont {Way},
  \citenamefont {Morgan}, \citenamefont {Clauset},\ and\ \citenamefont
  {Larremore}}]{way:misleading}%
  \BibitemOpen
  \bibfield  {author} {\bibinfo {author} {\bibfnamefont {Samuel~F}\
  \bibnamefont {Way}}, \bibinfo {author} {\bibfnamefont {Allison~C}\
  \bibnamefont {Morgan}}, \bibinfo {author} {\bibfnamefont {Aaron}\
  \bibnamefont {Clauset}}, \ and\ \bibinfo {author} {\bibfnamefont {Daniel~B}\
  \bibnamefont {Larremore}},\ }\bibfield  {title} {\enquote {\bibinfo {title}
  {The misleading narrative of the canonical faculty productivity
  trajectory},}\ }\href@noop {} {\bibfield  {journal} {\bibinfo  {journal}
  {Proc.\ Natl.\ Acad.\ Sci.\ USA}\ }\textbf {\bibinfo {volume} {114}},\
  \bibinfo {pages} {E9216--E9223} (\bibinfo {year} {2017})}\BibitemShut
  {NoStop}%
\bibitem [{\citenamefont {Groves}\ \emph {et~al.}(2011)\citenamefont {Groves},
  \citenamefont {Fowler~Jr}, \citenamefont {Couper}, \citenamefont {Lepkowski},
  \citenamefont {Singer},\ and\ \citenamefont {Tourangeau}}]{groves2011survey}%
  \BibitemOpen
  \bibfield  {author} {\bibinfo {author} {\bibfnamefont {Robert~M}\
  \bibnamefont {Groves}}, \bibinfo {author} {\bibfnamefont {Floyd~J}\
  \bibnamefont {Fowler~Jr}}, \bibinfo {author} {\bibfnamefont {Mick~P}\
  \bibnamefont {Couper}}, \bibinfo {author} {\bibfnamefont {James~M}\
  \bibnamefont {Lepkowski}}, \bibinfo {author} {\bibfnamefont {Eleanor}\
  \bibnamefont {Singer}}, \ and\ \bibinfo {author} {\bibfnamefont {Roger}\
  \bibnamefont {Tourangeau}},\ }\href@noop {} {\emph {\bibinfo {title} {Survey
  methodology}}},\ Vol.\ \bibinfo {volume} {561}\ (\bibinfo  {publisher} {John
  Wiley \& Sons},\ \bibinfo {year} {2011})\BibitemShut {NoStop}%
\bibitem [{\citenamefont {Weisberg}(2009)}]{weisberg2009total}%
  \BibitemOpen
  \bibfield  {author} {\bibinfo {author} {\bibfnamefont {Herbert~F}\
  \bibnamefont {Weisberg}},\ }\href@noop {} {\emph {\bibinfo {title} {The total
  survey error approach: A guide to the new science of survey research}}}\
  (\bibinfo  {publisher} {University of Chicago Press},\ \bibinfo {year}
  {2009})\BibitemShut {NoStop}%
\bibitem [{\citenamefont {Imai}\ \emph {et~al.}(2011)\citenamefont {Imai},
  \citenamefont {Keele}, \citenamefont {Tingley},\ and\ \citenamefont
  {Yamamoto}}]{imai2011unpacking}%
  \BibitemOpen
  \bibfield  {author} {\bibinfo {author} {\bibfnamefont {Kosuke}\ \bibnamefont
  {Imai}}, \bibinfo {author} {\bibfnamefont {Luke}\ \bibnamefont {Keele}},
  \bibinfo {author} {\bibfnamefont {Dustin}\ \bibnamefont {Tingley}}, \ and\
  \bibinfo {author} {\bibfnamefont {Teppei}\ \bibnamefont {Yamamoto}},\
  }\bibfield  {title} {\enquote {\bibinfo {title} {Unpacking the black box of
  causality: Learning about causal mechanisms from experimental and
  observational studies},}\ }\href@noop {} {\bibfield  {journal} {\bibinfo
  {journal} {American Political Science Review}\ }\textbf {\bibinfo {volume}
  {105}},\ \bibinfo {pages} {765--789} (\bibinfo {year} {2011})}\BibitemShut
  {NoStop}%
\bibitem [{\citenamefont {Menczer}(1997)}]{menczer:arachnid}%
  \BibitemOpen
  \bibfield  {author} {\bibinfo {author} {\bibfnamefont {Filippo}\ \bibnamefont
  {Menczer}},\ }\bibfield  {title} {\enquote {\bibinfo {title} {Arachnid:
  Adaptive retrieval agents choosing heuristic neighborhoods for information
  discovery},}\ }in\ \href@noop {} {\emph {\bibinfo {booktitle} {Proc.\ 14th
  Internat.\ Conf.\ Machine Learning}}}\ (\bibinfo {year} {1997})\ pp.\
  \bibinfo {pages} {227--235}\BibitemShut {NoStop}%
\bibitem [{\citenamefont {Mitchell}(2015)}]{mitchell:nell}%
  \BibitemOpen
  \bibfield  {author} {\bibinfo {author} {\bibfnamefont {Tom~M}\ \bibnamefont
  {Mitchell}},\ }\bibfield  {title} {\enquote {\bibinfo {title} {Never-ending
  learning},}\ }in\ \href@noop {} {\emph {\bibinfo {booktitle} {Proc.\ 29th
  Conf.\ Artificial Intelligence (AAAI)}}}\ (\bibinfo {year} {2015})\ pp.\
  \bibinfo {pages} {2302--2310}\BibitemShut {NoStop}%
\bibitem [{\citenamefont {McCallum}\ \emph {et~al.}(2000)\citenamefont
  {McCallum}, \citenamefont {Nigam}, \citenamefont {Rennie},\ and\
  \citenamefont {Seymore}}]{mccallum:automating}%
  \BibitemOpen
  \bibfield  {author} {\bibinfo {author} {\bibfnamefont {Andrew~Kachites}\
  \bibnamefont {McCallum}}, \bibinfo {author} {\bibfnamefont {Kamal}\
  \bibnamefont {Nigam}}, \bibinfo {author} {\bibfnamefont {Jason}\ \bibnamefont
  {Rennie}}, \ and\ \bibinfo {author} {\bibfnamefont {Kristie}\ \bibnamefont
  {Seymore}},\ }\bibfield  {title} {\enquote {\bibinfo {title} {Automating the
  construction of internet portals with machine learning},}\ }\href@noop {}
  {\bibfield  {journal} {\bibinfo  {journal} {Information Retrieval}\ }\textbf
  {\bibinfo {volume} {3}},\ \bibinfo {pages} {127--163} (\bibinfo {year}
  {2000})}\BibitemShut {NoStop}%
\bibitem [{\citenamefont {Cho}\ \emph {et~al.}(1998)\citenamefont {Cho},
  \citenamefont {Garcia-Molina},\ and\ \citenamefont {Page}}]{cho:efficient}%
  \BibitemOpen
  \bibfield  {author} {\bibinfo {author} {\bibfnamefont {Junghoo}\ \bibnamefont
  {Cho}}, \bibinfo {author} {\bibfnamefont {Hector}\ \bibnamefont
  {Garcia-Molina}}, \ and\ \bibinfo {author} {\bibfnamefont {Lawrence}\
  \bibnamefont {Page}},\ }\bibfield  {title} {\enquote {\bibinfo {title}
  {Efficient crawling through {URL} ordering},}\ }\href@noop {} {\bibfield
  {journal} {\bibinfo  {journal} {Computer Networks and ISDN Systems}\ }\textbf
  {\bibinfo {volume} {30}},\ \bibinfo {pages} {161--172} (\bibinfo {year}
  {1998})}\BibitemShut {NoStop}%
\bibitem [{\citenamefont {Menczer}\ \emph {et~al.}(2004)\citenamefont
  {Menczer}, \citenamefont {Pant},\ and\ \citenamefont
  {Srinivasan}}]{menczer:topical}%
  \BibitemOpen
  \bibfield  {author} {\bibinfo {author} {\bibfnamefont {Filippo}\ \bibnamefont
  {Menczer}}, \bibinfo {author} {\bibfnamefont {Gautam}\ \bibnamefont {Pant}},
  \ and\ \bibinfo {author} {\bibfnamefont {Padmini}\ \bibnamefont
  {Srinivasan}},\ }\bibfield  {title} {\enquote {\bibinfo {title} {Topical web
  crawlers: Evaluating adaptive algorithms},}\ }\href@noop {} {\bibfield
  {journal} {\bibinfo  {journal} {Transactions on Internet Technology}\
  }\textbf {\bibinfo {volume} {4}},\ \bibinfo {pages} {378--419} (\bibinfo
  {year} {2004})}\BibitemShut {NoStop}%
\bibitem [{\citenamefont {Pant}\ \emph {et~al.}(2004)\citenamefont {Pant},
  \citenamefont {Srinivasan},\ and\ \citenamefont {Menczer}}]{pant:crawling}%
  \BibitemOpen
  \bibfield  {author} {\bibinfo {author} {\bibfnamefont {Gautam}\ \bibnamefont
  {Pant}}, \bibinfo {author} {\bibfnamefont {Padmini}\ \bibnamefont
  {Srinivasan}}, \ and\ \bibinfo {author} {\bibfnamefont {Filippo}\
  \bibnamefont {Menczer}},\ }\bibfield  {title} {\enquote {\bibinfo {title}
  {Crawling the web},}\ }in\ \href@noop {} {\emph {\bibinfo {booktitle} {Web
  Dynamics}}},\ \bibinfo {editor} {edited by\ \bibinfo {editor} {\bibfnamefont
  {M}~\bibnamefont {Levene}}\ and\ \bibinfo {editor} {\bibfnamefont
  {A}~\bibnamefont {Poulovassilis}}}\ (\bibinfo  {publisher} {Springer},\
  \bibinfo {year} {2004})\ pp.\ \bibinfo {pages} {153--178}\BibitemShut
  {NoStop}%
\bibitem [{\citenamefont {Chakrabarti}\ \emph {et~al.}(1999)\citenamefont
  {Chakrabarti}, \citenamefont {van~den Berg},\ and\ \citenamefont
  {Dom}}]{chakrabarti:focused}%
  \BibitemOpen
  \bibfield  {author} {\bibinfo {author} {\bibfnamefont {Soumen}\ \bibnamefont
  {Chakrabarti}}, \bibinfo {author} {\bibfnamefont {Martin}\ \bibnamefont
  {van~den Berg}}, \ and\ \bibinfo {author} {\bibfnamefont {Byron}\
  \bibnamefont {Dom}},\ }\bibfield  {title} {\enquote {\bibinfo {title}
  {Focused crawling: A new approach to topic-specific web resource
  discovery},}\ }in\ \href {http://dl.acm.org/citation.cfm?id=313234.313121}
  {\emph {\bibinfo {booktitle} {Proc.\ 8th Internat.\ Conf.\ on World Wide Web
  (WWW)}}},\ \bibinfo {series and number} {WWW '99}\ (\bibinfo  {publisher}
  {Elsevier North-Holland, Inc.},\ \bibinfo {address} {New York, NY, USA},\
  \bibinfo {year} {1999})\ pp.\ \bibinfo {pages} {1623--1640}\BibitemShut
  {NoStop}%
\bibitem [{\citenamefont {Menczer}\ \emph {et~al.}(2001)\citenamefont
  {Menczer}, \citenamefont {Pant}, \citenamefont {Srinivasan},\ and\
  \citenamefont {Ruiz}}]{menczer:evaluating}%
  \BibitemOpen
  \bibfield  {author} {\bibinfo {author} {\bibfnamefont {Filippo}\ \bibnamefont
  {Menczer}}, \bibinfo {author} {\bibfnamefont {Gautam}\ \bibnamefont {Pant}},
  \bibinfo {author} {\bibfnamefont {Padmini}\ \bibnamefont {Srinivasan}}, \
  and\ \bibinfo {author} {\bibfnamefont {Miguel~E.}\ \bibnamefont {Ruiz}},\
  }\bibfield  {title} {\enquote {\bibinfo {title} {Evaluating topic-driven web
  crawlers},}\ }in\ \href {\doibase 10.1145/383952.383995} {\emph {\bibinfo
  {booktitle} {Proc.\ 24th Internat.\ ACM SIGIR Conf.\ on Research and
  Development in Information Retrieval}}},\ \bibinfo {series and number} {SIGIR
  '01}\ (\bibinfo  {publisher} {ACM},\ \bibinfo {address} {New York, NY, USA},\
  \bibinfo {year} {2001})\ pp.\ \bibinfo {pages} {241--249}\BibitemShut
  {NoStop}%
\bibitem [{\citenamefont {Butt}\ \emph {et~al.}(2015)\citenamefont {Butt},
  \citenamefont {Haller},\ and\ \citenamefont {Xie}}]{butt:taxonomy}%
  \BibitemOpen
  \bibfield  {author} {\bibinfo {author} {\bibfnamefont {Anila~Sahar}\
  \bibnamefont {Butt}}, \bibinfo {author} {\bibfnamefont {Armin}\ \bibnamefont
  {Haller}}, \ and\ \bibinfo {author} {\bibfnamefont {Lexing}\ \bibnamefont
  {Xie}},\ }\bibfield  {title} {\enquote {\bibinfo {title} {A taxonomy of
  semantic web data retrieval techniques},}\ }in\ \href@noop {} {\emph
  {\bibinfo {booktitle} {Proc.\ 8th Internat.\ Conf.\ Knowledge Capture}}}\
  (\bibinfo {year} {2015})\ p.~\bibinfo {pages} {9}\BibitemShut {NoStop}%
\bibitem [{\citenamefont {Xu}(2008)}]{xu2008gender}%
  \BibitemOpen
  \bibfield  {author} {\bibinfo {author} {\bibfnamefont {Yonghong~Jade}\
  \bibnamefont {Xu}},\ }\bibfield  {title} {\enquote {\bibinfo {title} {Gender
  disparity in stem disciplines: A study of faculty attrition and turnover
  intentions},}\ }\href@noop {} {\bibfield  {journal} {\bibinfo  {journal}
  {Research in Higher Education}\ }\textbf {\bibinfo {volume} {49}},\ \bibinfo
  {pages} {607--624} (\bibinfo {year} {2008})}\BibitemShut {NoStop}%
\bibitem [{\citenamefont {Kulis}\ \emph {et~al.}(2002)\citenamefont {Kulis},
  \citenamefont {Sicotte},\ and\ \citenamefont {Collins}}]{kulis2002more}%
  \BibitemOpen
  \bibfield  {author} {\bibinfo {author} {\bibfnamefont {Stephen}\ \bibnamefont
  {Kulis}}, \bibinfo {author} {\bibfnamefont {Diane}\ \bibnamefont {Sicotte}},
  \ and\ \bibinfo {author} {\bibfnamefont {Shawn}\ \bibnamefont {Collins}},\
  }\bibfield  {title} {\enquote {\bibinfo {title} {More than a pipeline
  problem: Labor supply constraints and gender stratification across academic
  science disciplines},}\ }\href@noop {} {\bibfield  {journal} {\bibinfo
  {journal} {Research in Higher Education}\ }\textbf {\bibinfo {volume} {43}},\
  \bibinfo {pages} {657--691} (\bibinfo {year} {2002})}\BibitemShut {NoStop}%
\bibitem [{\citenamefont {Pell}(1996)}]{pell1996fixing}%
  \BibitemOpen
  \bibfield  {author} {\bibinfo {author} {\bibfnamefont {Alice~N.}\
  \bibnamefont {Pell}},\ }\bibfield  {title} {\enquote {\bibinfo {title}
  {Fixing the leaky pipeline: women scientists in academia.}}\ }\href@noop {}
  {\bibfield  {journal} {\bibinfo  {journal} {J.\ Animal Sci}\ }\textbf
  {\bibinfo {volume} {74}},\ \bibinfo {pages} {2843--2848} (\bibinfo {year}
  {1996})}\BibitemShut {NoStop}%
\bibitem [{\citenamefont {Jadidi}\ \emph {et~al.}(2017)\citenamefont {Jadidi},
  \citenamefont {Karimi},\ and\ \citenamefont {Wagner}}]{jadidi2017gender}%
  \BibitemOpen
  \bibfield  {author} {\bibinfo {author} {\bibfnamefont {Mohsen}\ \bibnamefont
  {Jadidi}}, \bibinfo {author} {\bibfnamefont {Fariba}\ \bibnamefont {Karimi}},
  \ and\ \bibinfo {author} {\bibfnamefont {Claudia}\ \bibnamefont {Wagner}},\
  }\href@noop {} {\enquote {\bibinfo {title} {Gender disparities in science?
  {D}ropout, productivity, collaborations and success of male and female
  computer scientists},}\ } (\bibinfo {year} {2017}),\ \bibinfo {note}
  {preprint, {\tt arXiv:1704.05801}}\BibitemShut {NoStop}%
\bibitem [{\citenamefont {Rosvall}\ and\ \citenamefont
  {Bergstrom}(2008)}]{rosvall:map:2008}%
  \BibitemOpen
  \bibfield  {author} {\bibinfo {author} {\bibfnamefont {Martin}\ \bibnamefont
  {Rosvall}}\ and\ \bibinfo {author} {\bibfnamefont {Carl~T.}\ \bibnamefont
  {Bergstrom}},\ }\bibfield  {title} {\enquote {\bibinfo {title} {Maps of
  random walks on complex networks reveal community structure},}\ }\href@noop
  {} {\bibfield  {journal} {\bibinfo  {journal} {Proc.\ Natl.\ Acad.\ Sci.\
  USA}\ }\textbf {\bibinfo {volume} {105}},\ \bibinfo {pages} {1118--1123}
  (\bibinfo {year} {2008})}\BibitemShut {NoStop}%
\end{thebibliography}
\end{document}